\documentclass[aip,apl,twocolumn,reprint]{revtex4-1}

\usepackage{graphicx}
\usepackage{epstopdf}
\usepackage{amsmath}
\usepackage{natbib}

\begin{document}

\title{Full control of quadruple quantum dot circuit charge states in the single electron regime}

\author{M.~R.~Delbecq}
\email[]{matthieu.delbecq@riken.jp}
\author{T.~Nakajima}
\author{T.~Otsuka}
\author{S.~Amaha}
\affiliation{RIKEN, Center for Emergent Matter Science, 3-1 Wako-shi, Saitama 351-0198, Japan}
\author{J.~D.~Watson}
\affiliation{Department of Physics, Purdue University, West Lafayette, Indiana 47907, USA}
\affiliation{Birck Nanotechnology Center, Purdue University, West Lafayette, Indiana 47907, USA}
\author{M.~J.~Manfra}
\affiliation{Department of Physics, Purdue University, West Lafayette, Indiana 47907, USA}
\affiliation{Birck Nanotechnology Center, Purdue University, West Lafayette, Indiana 47907, USA}
\affiliation{School of Materials Engineering, Purdue University, West Lafayette, Indiana 47907, USA}
\affiliation{School of Electrical and Computer Engineering, Purdue University, West Lafayette, Indiana 47907, USA}
\author{S.~Tarucha}
\affiliation{RIKEN, Center for Emergent Matter Science, 3-1 Wako-shi, Saitama 351-0198, Japan}
\affiliation{Department of Applied Physics, University of Tokyo, 7-3-1 Hongo, Bunkyo-ku, Tokyo 113-8656, Japan}

\date{\today}

\begin{abstract}
We report the realization of an array of four tunnel coupled quantum dots in the single electron regime, which is the first required step toward a scalable solid state spin qubit architecture. We achieve an efficient tunability of the system but also find out that the conditions to realize spin blockade readout are not as straightforwardly obtained as for double and triple quantum dot circuits. We use a simple capacitive model of the series quadruple quantum dots circuit to investigate its complex charge state diagrams and are able to find the most suitable configurations for future Pauli spin blockade measurements. We then experimentally realize the corresponding charge states with a good agreement to our model.
\end{abstract}


\maketitle

Quantum dot (QD) circuits have demonstrated to be particularly good systems for studying electronic transport and for implementing solid state qubits. They notably offer the possibility to control the spin of confined single electrons to realize spin qubits \cite{Petta2005,Koppens2006}. These are especially attracting for quantum information processing because of their robustness to decoherence \cite{DiVincenzo1995,Hanson2007} which should allow to implement a full electron-spin based quantum computation scheme \cite{Loss98}. Among the various materials in which such QDs based spin qubits have been demonstrated, semiconductor heterostructures are considered candidates of choice because of the high tunability and readout techniques they offer. Lately, several experiments demonstrated the manipulation of two spin-1/2 qubits implemented in double QD (DQD) circuits as well as the realization of universal quantum one- and two- qubits gate operations \cite{Pioro08,Obata10,Brunner11}. An additional key feature of these semiconductor QD circuits is their potential for scalability \cite{Taylor05}.

Realization of a scalable architecture of semiconductor spin qubits is one of the remaining challenge that has to be overcome for implementing more complex algorithms. Steps toward this direction have been taken by experimentally realizing triple QD (TQD) circuits \cite{Medford2013,Braakman2013} or quadruple QD (QQD) circuits formed by two capacitively coupled DQDs \cite{Shulman2012}. In these systems however, the number of implemented qubits is still limited to one or two. A square-like configuration of tunnel coupled QQD device has also been demonstrated \cite{Thalineau2012}, in the single electron regime already. This particular configuration is however a priori less suitable for scalability than the series configuration discussed in this letter. Very recently, a tunnel coupled series-QQD device has been realized and studied in the multiple electrons regime \cite{Takakura2014}, showing a good control of the tunnel couplings and gate potentials to form the dots. In this letter, we show that by following the same architecture, we can reach the single electron regime for each QD, which is a mandatory condition for making four spin qubits. We furthermore demonstrate an efficient tunability of the system in this regime, with which we are able to realize the proper charge state configurations for spin blockade readout.

\begin{figure}
\includegraphics[width=8.5cm]{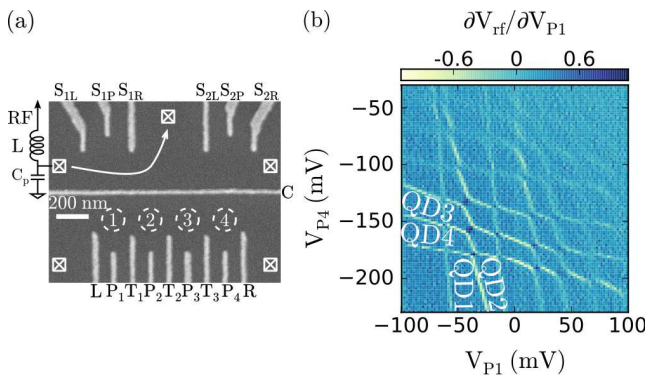}
\caption{\label{fig:1}(Color online) (a) Scanning electron microscopy (SEM) image of the QQD device.  The locations of the QDs are indicated by dashed white circles. The ohmic contacts are shown as crossed white boxes. Throughout this paper, the measurements are performed using the left charge sensor, as indicated by the white arrow, defined by the gates $\mathrm{S_{1L}}$, $\mathrm{S_{1P}}$ and $\mathrm{S_{1R}}$. Charge sensing is done by rf-reflectometry measurement. (b) Charge stability diagram of the QQD in the plane defined by plunger gates $\mathrm{P_1}$ and $\mathrm{P_4}$, with $V_{P2} = -125$ mV and $V_{P3}= -10$ mV. The color code is the derivative of the rf demodulated signal with respect to $V_{P1}$. The QQD empty state (0,0,0,0) is observed in the lower left corner.}
\end{figure}

The scanning electron microscopy image of the device studied here is shown in Fig. \ref{fig:1} (a). The quantum dots are formed in a 91 nm deep two-dimensional electron gas (2DEG) formed in a GaAs/AlGaAs heterostructure. The positions of the QDs is defined by a series of gates, L, $\mathrm{P_i}$, $\mathrm{T_j}$ and R and a long back wall gate C. The plunger gates $\mathrm{P_i}$ allow for controlling the $i$-th QD ($\mathrm{QD_i}$) energy while the tunnel gates $\mathrm{T_j}$ allow for controlling the tunnel coupling between the two adjacent $\mathrm{QD_j}$ and $\mathrm{QD_{j+1}}$. The device is prepared with a micro-magnet so as to fit future requirements for forming and manipulating spin qubits \cite{Obata10}. Two sets of three gates, visible on top of gate C, allow for forming QDs in multiple electrons regime for charge sensing purpose. Each of these charge sensors can be used as radio frequency (rf) charge sensor \cite{Reilly2007,Cassidy2007,Muller2007}. Throughout this paper, we use the rf-sensor on the left, whose resonant circuit frequency is $f=\mathrm{203.9\ MHz}$. All the measurements are performed at the base temperature of around 10 mK.

The voltage applied to gate C is $V_C =-450$ mV, below the pinch off voltage of the 2DEG ($V_{pinch-off}=-350$ mV). We can form the four QDs as indicated by white circles in Fig. \ref{fig:1} (a) by applying negative voltages to the other gates. Fig. \ref{fig:1} (b) shows the stability diagram obtained by modulating $V_{P1}$ and $V_{P4}$, the voltages respectively applied to plunger gates $\mathrm{P_1}$ and $\mathrm{P_4}$. The measured signal is the derivative of the rf demodulated signal amplitude with respect to $V_{P1}$, so as to display the charge transition lines of each QD. Four different slopes are identifiable, corresponding to the four different QDs. The stability diagram is in accordance with the device geometry as the slopes of each QD are directly related to the QD distance to the modulating plunger gates. Therefore we can assign the charge state $(N_1, N_2, N_3, N_4)$, with $N_i$ the number of electrons in $\mathrm{QD_i}$. We find that the single electron regime is already demonstrated in this stability diagram, as the charge state (0,0,0,0) is observed for $V_{P1} \leq - 50$ mV and $V_{P4} \leq - 170$ mV.

\begin{figure}
\includegraphics[width=8.5cm]{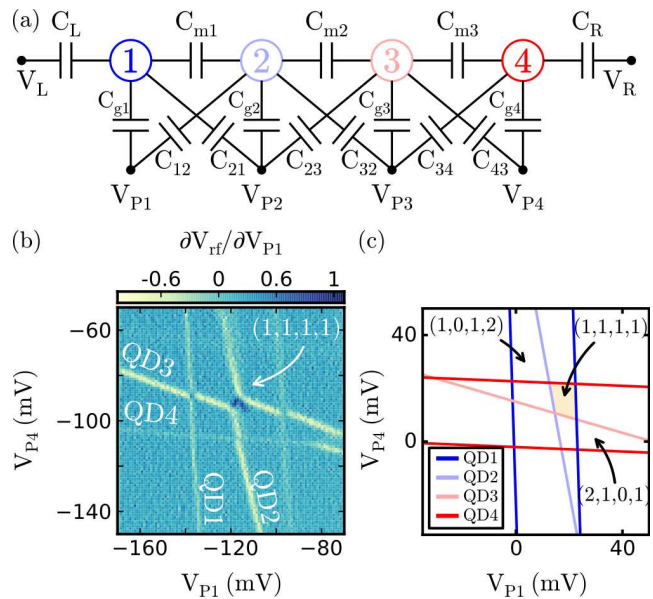}
\caption{\label{fig:2}(Color online) (a) Schematic of the QQD capacitive model used throughout this paper to calculate the charge stability diagrams. (b) Charge stability diagram close-up of the (1,1,1,1) region, in the plane defined by plunger gates $\mathrm{P_1}$ and $\mathrm{P_4}$, with $V_{P2} =-40$mV and $V_{P3}=-180$ mV. (c) Calculated transition lines using the model of (a) and $\mathrm{C_{gi} = C_{mi} = C_L = C_R = 10\ aF}$ and $\mathrm{C_{i\pm1,i} = 1\ aF}$ for $i = 1,2,3,4$.}
\end{figure}

We model our QQD device as shown in Fig. \ref{fig:2} (a). This purely classical model is an extension of the capacitive DQD model \cite{Wiel2002} to four QDs. We consider each $\mathrm{QD_i}$ to be capacitively coupled to its plunger gate via $\mathrm{C_{gi}}$, to the nearest neighbor plunger gate via $\mathrm{C_{i\pm1,i}}$ and to the nearest QD via the mutual capacitance $\mathrm{C_{mi}}$. QD1 and QD4 are also coupled to the left and right leads via $\mathrm{C_{L}}$ and $\mathrm{C_{R}}$, respectively. We therefore have 15 parameters to adjust in this model. However thanks to the symmetry of the device pattern and the slopes of the transition lines measured in the stability diagram, we can reproduce with good agreement the transition lines around the (1,1,1,1) region as shown in Fig. \ref{fig:2} (b) and (c) with $\mathrm{C_{gi} = C_{mi} = C_L = C_R = 10\ aF}$ and $\mathrm{C_{i\pm1,i} = 1\ aF}$ for $i = 1,2,3,4$. Note that we chose to only consider first neighbor plunger gate cross-capacitive coupling. This model already gives good qualitative and quantitative agreements with our data and as such is a good trade-off with more complete models requiring additional parameters.

We then need to consider the realization of spin readout. This mandatory feature for any purpose of quantum information manipulation is usually performed by Pauli spin blockade technique (PSB). It can be performed by DC measurements of the current in the biased regime \cite{Ono2002} or by pulse measurement techniques at high frequencies \cite{Petta2005}. In both of these schemes, it is necessary that two neighboring QDs have for adjacent charge states (2,0) (or (0,2)) and (1,1).

This scheme can in principle be extended to larger number of series QDs, by applying the pulsed PSB scheme to successive DQDs of the array. In the case of a TQD circuit, the PSB measurement would be done on the left DQD, $(2,0,1)\leftrightarrow (1,1,1)$, and right DQD $(1,0,2)\leftrightarrow (1,1,1)$, with the center QD being common \cite{Medford2013}. For QQDs, we are similarly looking for PSB conditions on the left DQD, $(2,0,1,1)\leftrightarrow (1,1,1,1)$, and right DQD, $(1,1,0,2)\leftrightarrow (1,1,1,1)$. The boundaries of the charge state (1,1,1,1) is defined by a transition line of each QDs. However, to meet the conditions of PSB for one DQD, we need to have the transition lines of the corresponding two QDs to cross and form one corner of the (1,1,1,1) region. Fig. \ref{fig:2} (b) and (c) show that none of the required conditions for PSB are met as neither the (2,0,1,1) nor the (1,1,0,2) charge states are adjacent to the (1,1,1,1) region. In this diagram, in order to achieve the PSB condition on the left DQD, one has to push the transition line of QD2 towards more positive $V_{P1}$ so that it crosses the second transition line of QD1 above the transition line of QD3 along $V_{P4}$. Similarly, to obtain the PSB condition on the right DQD, one has to  push the transition line of QD3 towards more positive $V_{P4}$ so that it crosses the second transition line of QD4 above the transition line of QD2 along $V_{P1}$. Following this procedure, and with the help of the QQD capacitive model, we find that there exists no configuration in which the left and right DQDs PSB conditions can be met on a single stability diagram defined by two plunger gates. This fact sets a clear gap with TQD devices in the search of a scalable architecture, as both PSB conditions can be found on the same stability diagram for TQD. It implies that more complex manipulations of the QQD system are necessary to fully operate and measure the spin state of each QDs.

\begin{figure}
\includegraphics[width=8.5cm]{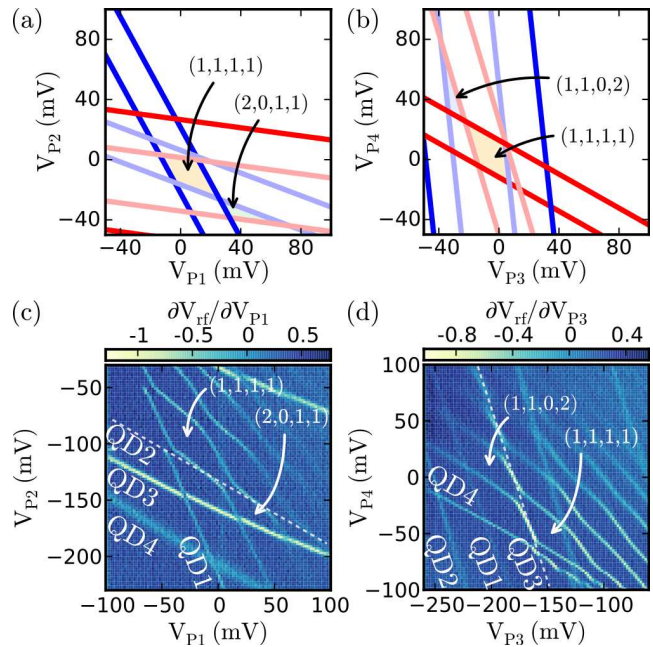}
\caption{\label{fig:3}(Color online) Suitable regions for spin blockade measurements. (a) Calculated stability diagram in the plane defined by plunger gates $\mathrm{P_1}$ and $\mathrm{P_2}$, allowing for PSB condition in the left DQD. (b) Calculated stability diagram in the plane defined by plunger gates $\mathrm{P_3}$ and $\mathrm{P_4}$, allowing for PSB condition in the right DQD. (c) Charge stability diagram in the plane defined by plunger gates $\mathrm{P_1}$ and $\mathrm{P_2}$, with $V_{P3} =-60$ mV and $V_{P4}=-10$ mV. The two regions (1,1,1,1) and (2,0,1,1) between which the spin blockade measurements on the left DQD can be performed are shown by white arrows. (d) Charge stability diagram in the plane defined by plunger gates $\mathrm{P_3}$ and $\mathrm{P_4}$, with $V_{P1} =-335$ mV and $V_{P2}=-190$ mV. The two regions (1,1,1,1) and (1,1,0,2) between which the spin blockade measurements on the right DQD can be performed are shown by white arrows. The color code of the transition lines in (a) and (b) corresponds to the one of Fig. \ref{fig:2} (b).}
\end{figure}

Up to now, we only considered the stability diagram of the QQD in the $(V_{P1},V_{P4})$ plane. We however have a set of four available plunger gates to explore the complete manifold of the QQD charge states. Within this picture, each charge state region is 4-dimensional and can be explored along the axes $V_{P1}$, $V_{P2}$, $V_{P3}$ and $V_{P4}$. Our simple capacitive model then reveals especially useful to explore the charge states space along any combination of these axes. 

Fig. \ref{fig:3} (a) and (b) show the calculation of the stability diagrams (where the first two electrons of each QDs are considered) in the planes $(V_{P1},V_{P2})$ and $(V_{P3},V_{P4})$ respectively. The PSB conditions for each DQD is naturally found in each respective plane. The corresponding stability diagrams measurements are shown in Fig. \ref{fig:3} (c) and (d). Fig. \ref{fig:3} (a) (Fig. \ref{fig:3} (b)) shows that we should expect the transition lines of QDs 2, 3 and 4 (1, 2 and 3) to have similar slopes, little influenced by $V_{P1}$ ($V_{P4}$). The transition line spacing $\Delta V_i$ between two consecutive charge states of each $\mathrm{QD_i}$ is also directly related to the distance of each $\mathrm{QD_i}$ to the driving plunger gates \cite{Note1}, respectively giving $\Delta V_2 \leq \Delta V_3 \leq \Delta V_4$ and $\Delta V_3 \leq \Delta V_2 \leq \Delta V_1$. These features are well observed in Fig. \ref{fig:3} (c) and (d), confirming the agreement of the capacitive model of Fig. \ref{fig:2} (a) with our device. The comparison between theory and experiment allows us to clearly identify each QD transition line and find out the PSB conditions $(2,0,1,1)\leftrightarrow (1,1,1,1)$ in the plane $(V_{P1},V_{P2})$ and $(1,1,0,2)\leftrightarrow (1,1,1,1)$ in the plane $(V_{P3},V_{P4})$ as depicted by the white arrows.

An additional advantage of this scheme is that the transition lines of the two QDs where the PSB condition is met form a standard DQD honeycomb pattern in the corresponding plunger gate voltage plane. It provides better clarity for manipulating the left and right DQDs and allows for direct implementation of the standard DQD PSB schemes. This should particularly help limiting the modulation of gates potential and tunnel coupling of the two other QDs, that could give rise to spurious effects such as states mixing.

Ideally, one can first find the (1,1,1,1) region in the $(V_{P1},V_{P4})$ plane. Then, from that charge state, the charge stability diagram is explored in the other planes. One would then readily find the PSB conditions for each DQD with the minimum number of gate operations.

We thank J. Medford, J. Beil, F. Kuemmeth, C. M. Marcus, J. I. Colless and D. J. Reilly for helpful technical discussions. This work was supported by the Funding Program for World-Leading Innovative R\&D on Science and Technology (FIRST), the Grant-in-Aid for Young Scientists B from the Japan Society for the Promotion of Science, the IARPA project ``Multi-Qubit Coherent Operations'' through Copenhagen University, the Toyota Physical and Chemical Research Institute Scholars and the RIKEN Incentive Research Project.

\end{document}